\documentstyle[twocolumn,pre,aps,epsf]{revtex}

\newcommand{\be}{\begin{equation}}
\newcommand{\ee}{\end{equation}}
\newcommand{\notpar}{\not{}\hspace{-0.43em}\parallel}
\def\Ru{Sr$_2$RuO$_4$}
\def\Tc{{T_{\rm c}}}

\begin{document}
\draft
\twocolumn[\hsize\textwidth\columnwidth\hsize\csname @twocolumnfalse\endcsname

\title{Mechanism of spin-triplet superconductivity in Sr$_2$RuO$_4$}
\author{Masatoshi Sato and Mahito Kohmoto }
\address{Institute for Solid State Physics,
University of Tokyo, 7-22-1 Roppongi, Minato-ku, Tokyo, Japan}

\maketitle
\begin{abstract}
The unique Fermi surfaces and their nesting properties of \Ru ~are considered.
The existence of  unconventional superconductivity is shown microscopically,
for the first time, from the magnetic interactions (due to nesting) and the
phonon-mediated interactions.
The odd-parity superconductivity is favored in the $\alpha$ and $\beta$
sheets of the Fermi surface, and the various superconductivities are
possible in the $\gamma$ sheet.
There are a number of possible odd-parity gaps, which include the
gaps with nodes, the breaking of time-reversal symmetry and
$\vec{d}\parallel \hat{z}$.
\end{abstract}
\pacs{ 74.20.-z, 74.20.Fg}

]

\narrowtext
The nature of superconductivity discovered in \Ru \cite{maeno} has been the
subject of intense theoretical and experimental activity. Although \Ru~
has the same layered perovskite structure as La$_2$CuO$_4$, the
prototype of the high $\Tc$ cuprates superconductors, the electronic
structures are very different and the nature of superconductivity seems
to be totally different.

The normal state in \Ru ~is characterized as essentially a Fermi liquid
below $50$K.
The resistivities in all directions show $T^2$ behavior for $T \leq 50$K. The
effective mass is about $3 \sim 4m_{ {\rm electron}}$ and the susceptibility
is also about $3 \sim 4\chi_0$ where $\chi_0$ is the Pauli spin
susceptibility.
In contrast to the
conventional normal state (below $50$K), there are considerable experimental
evidences that the superconducting state (below about $1.5$K) is
unconventional.
The nuclear quadrupole resonance(NQR) does not show the Hebel-Slichter
peak\cite{ishida}. The transition temperature is very sensitive to non-magnetic
impurities\cite{mac}. The $^{17}$O NMR Knight shift shows that the spin
susceptibility has no change across $\Tc$ but stays just the same as in the
normal state for the magnetic field parallel to the \Ru ~plane
\cite{nmr}.
In addition, spontaneous appearance of an internal magnetic field below
the transition temperature is reported by muon spin rotation
measurements ($\mu SR$) \cite{muon}.

Shortly after the discovery of the superconductivity in \Ru, it was proposed
that the odd-parity(spin -triplet) Cooper pairs are formed in the
superconducting state in analogy with $^3$He \cite{sig}.
The existence of ferromagnetic interaction is crucial in this proposal.
In stead, incommensurate antiferromagnetic(AF) fluctuations were found
by inelastic neutron scattering experiment \cite{sidis}.
Ferromagnetic interaction is almost negligible compared with AF one.

Earlier specific heat\cite{nishi} and NQR measurements\cite{ishida} show a
large residual density of states (DOS), $50 \sim 60$\% of DOS of the
normal state, in the superconducting phase.
A possible explanation, so called, {\em orbital dependent
superconductivity} was proposed \cite{agter}.
Since four 4d electrons in Ru$^{4+}$
partially fill the $t_{2g}$ band, the relevant orbitals are $d_{xy}$,
$d_{xz}$ and $d_{yz}$ which determine the electronic bands.
The gap of of one class of bands is substantially smaller than that of
the other class of band.
The presence of gapless excitations for temperatures greater than the
smaller gap would account for the residual DOS.
Recent specific measurements on high quality compounds, however, suggest the
absence of residual DOS\cite{nishi2}.

Sigrist et al. \cite{sig2} proposed the following order parameter
which is claimed to be compatible
with most of the present experimental data,
\begin{eqnarray}
\vec{d}= \hat{z}(k_x \pm ik_y)
\label{d1},
\end{eqnarray}
where $\hat{z}$ is parallel to the $\hat{c}$ axis and the gap is described
as the tensor represented by $\vec{d}$ as
$
\Delta(k) = i(\vec{d}(k)\cdot\vec{\sigma})\sigma_y,
$
where $\vec{\sigma}$ is the Pauli matrix \cite{leggett}.
Notice that the direction of the order parameter is frozen
along the $\hat{c}$ direction due to the crystal field and there is a full
gap on the whole Fermi surface.
The experiment of Josephson coupling between In and \Ru ~also supports
the gap directing $\hat{c}$ axis \cite{jin}.
Based on the gap (\ref{d1}), the effects of impurity scattering\cite{maki},
spin wave excitations\cite{kee} and
collective modes and sound propagation\cite{kim} are studied
theoretically.

However, the recent experiments on high quality compounds of Ru
NQR\cite{nqr} and specific heat\cite{nishi2} strongly suggest the existence of
nodes and the absence of residual DOS.
It is  proposed a number of  gap order parameters
consistent with the existence of nodes phenomenologically \cite{hasegawa}.
See also Ref.\cite{miyake}.
\begin{figure}
\centerline{\epsfxsize=5cm\epsfbox{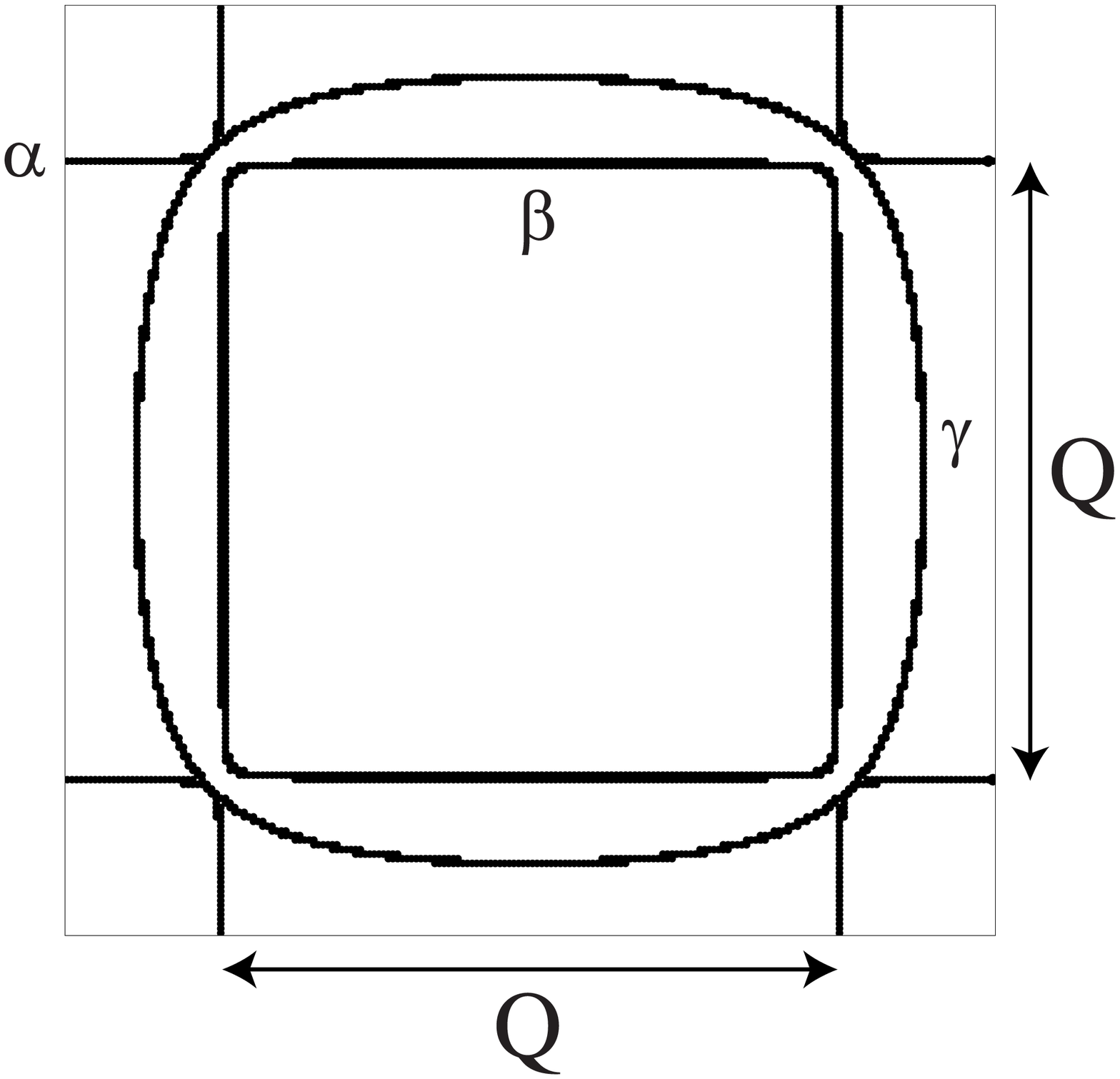}}
\caption{The Fermi surfaces of \Ru}
\label{fs:fig}
\end{figure}

Details of the Fermi surface have been observed by quantum oscillations
\cite{mac2}.
The Fermi surface consists of three sheets, which is
consistent with the electronic band calculations \cite{band}.
The Fermi sheets are labeled by $\alpha$, $\beta$, and $\gamma$.
See Fig. \ref{fs:fig}.
While the $\gamma$ sheet of the Fermi surface can be attributed solely
to $d_{xy}$ Wannier function, the $\alpha$ and $\beta$ sheets are due to
the hybridization of the $d_{xz}$ and $d_{yz}$ Wannier
functions.
The $\gamma$ band is quasi-isotropic
two-dimensional, on the other hand
the $\alpha$ and $\beta$ sheets are quasi-one dimensional which can be
visualized as a set of parallel planes separated by $Q = 4\pi/3$ running
both $k_x$ and $k_y$ directions.
Therefore it is natural to expect a sizable nesting effects at the wave
vectors $(\pm Q, k_y,k_z)$ and ($ k_x, \pm Q, k_z$).
The nesting vectors $( \pm Q, \pm Q, k_z)$ and $(\mp Q, \pm Q, k_z)$
have the maximum effects since they connect the one dimensional Fermi surfaces
in both directions. Collective modes in the spin dynamics are studied based on
these nesting effects
\cite{kee2}.
In fact the neutron scattering experiment\cite{sidis} shows peaks at
$(\pm 0.6\pi,\pm 0.6\pi)$ close to the nesting vectors (up to $(\pm
2\pi,\pm 2\pi)$) \cite{ms}.

\noindent
{\it -- Band structure and the pairing interactions}\\
We denote the annihilation operators for electrons in the
three $4d$-$t_{2g}$ orbitals $d_{xz}$, $d_{yz}$ and $d_{xy}$ of Ru-ions as
$a_{k,s}$, $b_{k,s}$ and $c_{k,s}$.
The kinetic term of the $\alpha$ and $\beta$ bands is given by
\begin{eqnarray}
H^{\alpha\beta}_{\rm kin}
&=&\sum_{k,s}(\varepsilon^{\alpha\beta} (k_x)-\mu)a_{k,s}^{\dagger}a_{k,s}
\nonumber\\
&+&\sum_{k,s}(\varepsilon^{\alpha\beta}(k_y)-\mu)b_{k,s}^{\dagger}b_{k,s}
\nonumber\\
&+&\sum_{k,s}t(k)a^{\dagger}_{k,s}b_{k,s}
+\sum_{k,s}t^{*}(k)b^{\dagger}_{k,s}a_{k,s},
\end{eqnarray}
and that of the $\gamma$ band is given by
\begin{eqnarray}
H^{\gamma}_{\rm kin}
=\sum_{k,s}(\varepsilon^{\gamma}(k_x,k_y)-\mu)c_{k,s}^{\dagger}c_{k,s}.
\end{eqnarray}
The $\alpha$ and $\beta$ bands are quasi one-dimensional and the
Fermi surfaces are well-approximated by four sheets $k_x\sim \pm Q/2$
and $k_y\sim\pm Q/2$.
The mixing coefficient $t(k)$ can be neglected except around $k_x=\pm k_y$.
The $\gamma$ band is quasi-isotropic two-dimensional one.
The mixing term between the $\alpha$ or $\beta$ band and the $\gamma$ one is
suppressed by the reflection symmetry of $z$.

The nesting of the $\alpha$ and $\beta$ bands leads to the following AF
fluctuations:
\begin{eqnarray}
&&
H^{\alpha\beta}_{\rm AF}
\nonumber\\
=&&\sum_{k,q,s_i}
J^{\alpha\beta}_{\perp}(q_x)
(\sigma_{z})_{s_1s_3}\cdot(\sigma_{z})_{s_2s_4}
a^{\dagger}_{-k,s_1} a^{\dagger}_{k,s_2} a_{-k+q,s_3} a_{k-q,s_4}
\nonumber\\
+&&\sum_{k,q,s_i}
J^{\alpha\beta}_{\perp}(q_y)
(\sigma_{z})_{s_1s_3}\cdot(\sigma_{z})_{s_2s_4}
b^{\dagger}_{-k,s_1} b^{\dagger}_{k,s_2} b_{-k+q,s_3} b_{k-q,s_4}
\nonumber\\
+&&\sum_{k,q,s_i}
J^{\alpha\beta}_{\parallel}(q_x)
\left[
(\sigma_{x})_{s_1s_3}\cdot(\sigma_{x})_{s_2s_4}
+(\sigma_{y})_{s_1s_3}\cdot(\sigma_{y})_{s_2s_4}
\right]
\nonumber\\
&&\times
a^{\dagger}_{-k,s_1} a^{\dagger}_{k,s_2} a_{-k+q,s_3} a_{k-q,s_4}
\nonumber\\
+&&\sum_{k,q,s_i}
J^{\alpha\beta}_{\parallel}(q_y)
\left[
(\sigma_{x})_{s_1s_3}\cdot(\sigma_{x})_{s_2s_4}
+(\sigma_{y})_{s_1s_3}\cdot(\sigma_{y})_{s_2s_4}
\right]
\nonumber\\
&&\times
b^{\dagger}_{-k,s_1} b^{\dagger}_{k,s_2} b_{-k+q,s_3} b_{k-q,s_4},
\end{eqnarray}
where $J^{\alpha\beta}_{\perp}(q_i)>0$ and
$J^{\alpha\beta}_{\parallel}(q_i)>0$ have peaks at $q_i\sim \pm Q$.
In general,
$J^{\alpha\beta}_{\perp}(q_i) \neq J^{\alpha\beta}_{\parallel}(q_i)$.
The magnetic field generated by the AF fluctuations above
induces the interaction $H^{\gamma}_{\rm AF}$ in the $\gamma$ band:
\begin{eqnarray}
&&H^{\gamma}_{\rm AF}
\nonumber\\
=&&\sum_{k,q,s_i}
J_{\perp}^{\gamma}
(q_x,q_y)(\sigma_z)_{s_1s_3}\cdot(\sigma_z)_{s_2s_4}
c^{\dagger}_{-k,s_1} c^{\dagger}_{k,s_2} c_{-k+q,s_3} c_{k-q,s_3}
\nonumber\\
+&&\sum_{k,q,s_i}
J_{\parallel}^{\gamma}(q_x,q_y)
\left[
(\sigma_x)_{s_1s_3}\cdot(\sigma_x)_{s_2s_4}
+(\sigma_y)_{s_1s_3}\cdot(\sigma_y)_{s_2s_4}
\right]
\nonumber\\
&&\times
c^{\dagger}_{-k,s_1} c^{\dagger}_{k,s_2} c_{-k+q,s_3} c_{k-q,s_3},
\end{eqnarray}
where $J^{\gamma}_{\perp}(q_x,q_y)>0$ and $J^{\gamma}_{\parallel}(q_x,q_y)>0$
have peaks at $(q_x,q_y)=(\pm Q,\pm Q)$ and $(q_x,q_y)=(\mp Q, \pm Q)$.

In addition to the AF fluctuations, we consider the electron-phonon
interaction.
Due to the low dimensionality of the system, this interaction is
weakly screened.
We assume that the phonon-mediated interaction
for the $\alpha$ and $\beta$ bands $H^{\alpha\beta}_{\rm ph}$ is
quasi-one dimensional,
\begin{eqnarray}
H^{\alpha\beta}_{\rm ph}
&=&\sum_{k,k',s,s'} f^{\alpha\beta}(q_x)
a^{\dagger}_{-k,s} a^{\dagger}_{k,s'} a_{-k+q,s} a_{k-q,s'}
\nonumber\\
&+&\sum_{k,k',s,s'} f^{\alpha\beta}(q_y)
b^{\dagger}_{-k,s} b^{\dagger}_{k,s'} b_{-k+q,s} b_{k-q,s'},
\end{eqnarray}
where $f^{\alpha\beta}(q_i)>0$ has a peak at $q_i=0$.
For the $\gamma$ band, the phonon-mediated interaction we consider is
\begin{eqnarray}
H^{\gamma}_{\rm ph}
=\sum_{k,k',s,s'} f^{\gamma}(q_x,q_y)
c^{\dagger}_{-k,s} c^{\dagger}_{k,s'} c_{-k+q,s} c_{k-q,s'}
\end{eqnarray}
where $f^{\gamma}(q_x,q_y)>0$ has a peak at $(q_x,q_y)=(0,0)$.

\noindent
{\it -- Superconductivity in the $\alpha$ and $\beta$ bands}\\
First, we examine the superconductivity in the $\alpha$ and $\beta$
bands.
As we will show immediately, the odd-parity superconductivity is
realized due to the quasi-one dimensionality \cite{KS}.
For simplicity, we put
$f^{\alpha\beta}(q_i)\sim f^{\alpha\beta}(0)\delta_{q_i,0}$,
$J^{\alpha\beta}_{\perp}(q_i)
\sim J^{\alpha\beta}_{\perp}(Q)\delta_{q_i,\pm Q}$ and
$J^{\alpha\beta}_{\parallel}(q_i)
\sim J^{\alpha\beta}_{\parallel}(Q)\delta_{q_i,\pm Q}$ in the following.

In the lowest order approximation, we neglect the mixing term $t(k)$, so the
gap equation is separated for $a_{k,s}$ and $b_{k,s}$.
For $a_{k,s}$ electrons, the gap $\Delta^{(a)}$ is defined by
\begin{eqnarray}
\Delta^{(a)}_{s_2,s_1}(k)
=-\sum_{k',s_3,s_4} V_{s_1,s_2,s_3,s_4}(k_x,k'_x)a_{k',s_3} a_{-k',s_3},
\end{eqnarray}
where
\begin{eqnarray}
&&V_{s_1,s_2,s_3,s_4}(k_x,k_x')\nonumber\\
&&=f^{\alpha\beta}(k_x+k_x')\delta_{s_1s_3}\delta_{s_2s_4}
-f^{\alpha\beta}(k_x-k_x')\delta_{s_1s_4}\delta_{s_2s_3}
\nonumber\\
&&+J^{\alpha\beta}_{\perp}(k_x+k_x')
(\sigma_z)_{s_1s_3}\cdot(\sigma_z)_{s_2s_4}
\nonumber\\
&&-J^{\alpha\beta}_{\perp}(k_x-k_x')
(\sigma_z)_{s_1s_4}\cdot(\sigma_z)_{s_2s_3}.
\nonumber\\
&&+J^{\alpha\beta}_{\parallel}(k_x+k_x')
\left[
(\sigma_x)_{s_1s_3}\cdot(\sigma_x)_{s_2s_4}
+(\sigma_y)_{s_1s_3}\cdot(\sigma_y)_{s_2s_4}
\right]
\nonumber\\
&&-J^{\alpha\beta}_{\parallel}(k_x-k_x')
\left[
(\sigma_x)_{s_1s_4}\cdot(\sigma_x)_{s_2s_3}
+(\sigma_y)_{s_1s_4}\cdot(\sigma_y)_{s_2s_3}
\right].
\end{eqnarray}
If the gap is unitary (as we assume), the gap equation becomes
\begin{eqnarray}
\Delta^{(a)}_{s_2,s_1}(k)
&&=-\sum_{k',s_3,s_4}V_{s_1,s_2,s_3,s_4}(k_x,k_x')
\nonumber\\
&&\hspace{5ex}\times
\frac{\Delta_{s_3,s_4}^{(a)}(k')}{2 E^{(a)}_{k'}}
\tanh\left(\frac{\beta E^{(a)}_{k'}}{2}\right),
\end{eqnarray}
where $E^{(a)}_{k}=\sqrt{\epsilon^{\alpha\beta}(k_x)
+{\rm tr}(\Delta^{(a)}\Delta^{(a)\dagger})/2}$.
The gap $\Delta^{(a)}$ is parameterized as\cite{leggett}
\begin{eqnarray}
\Delta^{(a)}(k)
=
\left\{
\begin{array}{ll}
i\psi^{(a)}(k)\sigma_y & \mbox{for even-parity gap}\\
i(\vec{d}^{(a)}(k)\cdot\vec{\sigma})\sigma_y & \mbox{for odd-parity gap}
\end{array}
\right.,
\end{eqnarray}
and we obtain the solutions as
$\psi^{(a)}|_{k_x\sim\pm Q/2}={\rm const.}$
for even-parity gap and
$\vec{d}^{(a)}|_{k_x\sim\pm Q/2}=\pm{\rm const.}$ for odd-parity gap.
For these solutions, the critical temperature is
$
k_{\rm B}T_{\rm c}=1.13\hbar\omega_{\rm D}e^{-1/N(\mu)\lambda},
$
where $\omega_{\rm D}$ is the cut-off for the interactions,
$N(\mu)$ is the DOS at the Fermi surface for $a_{k,s}$ electrons and
$\lambda$ is
\begin{eqnarray}
\lambda=
\left\{
\begin{array}{ll}
f^{\alpha\beta}(0)-J^{\alpha\beta}_{\perp}(Q)-2J^{\alpha\beta}_{\parallel}(Q)
& \mbox{for even-parity gap}\\
f^{\alpha\beta}(0)+J^{\alpha\beta}_{\perp}(Q)-2J^{\alpha\beta}_{\parallel}(Q)
&  \mbox {for $\vec{d}^{(a)}\parallel \vec{z}$}\\
f^{\alpha\beta}(0)-J^{\alpha\beta}_{\perp}(Q)
&  \mbox {for $\vec{d}^{(a)}\perp \vec{z}$}
\end{array}
\right.
.
\end{eqnarray}
Therefore, the odd-parity superconductivity is realized in $a_{k,s}$,
and $\vec{d}^{(a)}$ vector becomes
\begin{eqnarray}
\vec{d}^{(a)}|_{ky\sim Q/2}=-\vec{d}^{(a)}|_{ky\sim-Q/2}={\rm const}.
\label{dvec},
\end{eqnarray}
and
\begin{eqnarray}
\begin{array}{ll}
\vec{d}^{(a)}\parallel \hat{z}
&\mbox{if  $J^{\alpha\beta}_{\perp}(Q) >J^{\alpha\beta}_{\parallel}(Q)$ }\\
\vec{d}^{(a)}\perp\hat{z}
&\mbox{if $J^{\alpha\beta}_{\perp}(Q)<J^{\alpha\beta}_{\parallel}(Q)$ }
\label{dvec2}
\end{array}.
\end{eqnarray}
Note that if $J^{\alpha\beta}_{\perp}(Q)$ is large enough, the
superconductivity is realized even when $f^{\alpha\beta}(0)=0$.

In a similar manner, the superconductivity in $b_{k,s}$ is
odd-parity and characterized by (\ref {dvec}) and (\ref {dvec2})~ when  every
$a$'s, $k_x$ and $k_y$ are replaced by $b$'s, $k_y$ and $k_x$.
The critical temperature is same as
$a_{k,s}$.
Due to $D_{4h}$ symmetry, it holds that $|\vec{d}^{(a)}|=|\vec{d}^{(b)}|$.

The mixing term $t(k)$ affects the gaps $\vec{d}^{(a)}$ and
$\vec{d}^{(b)}$ at $k=\pm K\equiv(\pm Q/2,\pm Q/2,k_z)$ and
$k=\pm\tilde{K}\equiv(\mp Q/2,\pm Q/2,k_z)$.
If $t(k)$ is large enough, the Fermi surface at
$k=\pm K$ and $k=\pm\tilde{K}$
is largely deformed.
Because of the weakness of the screening,
the attractive force for the electron at those points decreases.
This causes nodes at those points.
The relative phase of $\vec{d}^{(a)}$ and $\vec{d}^{(b)}$ is not
determined in this case, so the time-reversal symmetry is broken in
general.

\begin{figure}
\centerline{\epsfxsize=5cm\epsfbox{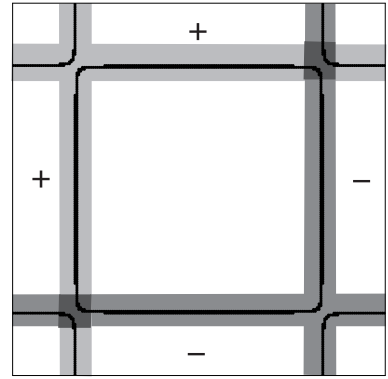}}
\caption{Superconductivity in the $\alpha$ and $\beta$ bands}
\label{albe:fig}
\end{figure}

If $t(k)$ is small, $\vec{d}^{(a)}$ and $\vec{d}^{(b)}$ do not disappear
at $k=\pm K$ and $k=\pm\tilde{K}$.
In this case, the relative phase of $\vec{d}^{(a)}$ and $\vec{d}^{(b)}$
is determined.
Since $t(k)$ at $k_x=\pm k_y$ is real, it can be easily shown that
$\arg\vec{d}^{(a)}=\arg\vec{d}^{(b)}$ or
$\arg\vec{d}^{(a)}=\arg\vec{d}^{(b)}+\pi$ at $k=\pm K$ and $k=\pm\tilde{K}$.
The superconductivity in this case is illustrated in Fig. \ref{albe:fig}.
The existence of nodes in the small $t(k)$ case depends on the symmetry
of the system.
If the system is invariant under the following transformation,
\begin{eqnarray}
\begin{array}{ll}
a_{k,s}\rightarrow b_{k,s}, \, b_{k,s}\rightarrow a_{k,s} &
\mbox{for $k_x=\pm k_y$}\\
a_{k,s}\rightarrow a_{k,s}, \, b_{k,s}\rightarrow b_{k,s} &
\mbox{for $k_x\neq \pm k_y$}
\end{array}
,
\label{nodesym:eqn}
\end{eqnarray}
the following pairing interaction is added.
\begin{eqnarray}
\Delta H^{\alpha\beta}_{\rm int}
=&&-\sum_{s,s'}
\Delta^{(a)*}_{s s'}(K)b_{K,s}b_{-K,s}+(\mbox{h.c.})
\nonumber\\
&&-\sum_{s,s'}\Delta^{(a)*}_{s s'}(\tilde{K})b_{\tilde{K},s}b_{-\tilde{K},s}
+(\mbox{h.c.})
\nonumber\\
&&+(a\leftrightarrow b).
\end{eqnarray}
This interaction changes the spectrum at $k=\pm K$ and $k=\pm\tilde{K}$.
If $\vec{d}^{(a)}\parallel\vec{d}^{(b)}$ holds,
nodes exists at either $k=\pm K$ or $k=\pm\tilde{K}$.
If the system does not have the symmetry above
or $\vec{d}^{(a)}\notpar\vec{d}^{(b)}$, no
node exists in the gap.
\noindent
{\it -- Superconductivity in the $\gamma$ band}\\
Next, we examine the superconductivity in the $\gamma$ band based on the
phonon-mediated interaction and  AF fluctuations.
In contrast to the $\alpha$ and $\beta$ bands, both  even-parity
 and odd-parity superconductivity  can be realized.
If $J^{\gamma}_{\perp}(q_x,q_y)$ is large enough, the odd-parity
superconductivity is also realized in the $\gamma$ band, but if not, the
various superconductivity can be realized.
To show this, we introduce the following AF interaction and
phonon-mediated interaction:
\begin{eqnarray}
&&J^{\gamma}_{{\perp \atopwithdelims\{\}\parallel}}(q_x,q_y)
\nonumber\\
&&=\sum_{Q_{x \atopwithdelims\{\}y}=\pm 4\pi/3}
\frac{g_{{\perp \atopwithdelims\{\} \parallel}}}
{(q_x-Q_x)^2+(q_y-Q_y)^2+Q_{\rm AF}^2},
\end{eqnarray}
and
\begin{eqnarray}
f^{\gamma}(q_x,q_y)=\frac{g_{\rm ph}}{q_x^2+q_y^2+Q_{\rm ph}^2},
\end{eqnarray}
where $g_{{\perp \atopwithdelims\{\}\parallel}}$, $g_{\rm ph}$,
$Q_{\rm ph}$ are constants.
{From} the experimental data in \cite{sidis},
we take $Q_{\rm AF}=0.0797\times 2\pi$.

\begin{figure}
\centerline{\epsfxsize=7cm\epsfbox{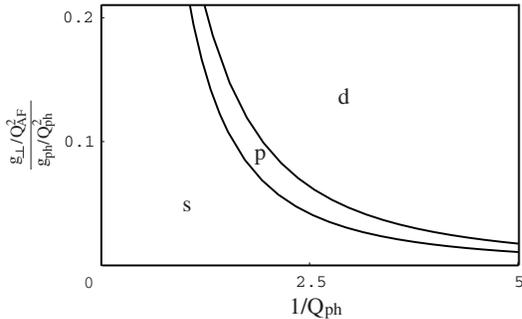}}
\caption{Superconductivity of the $\gamma$ band in the case where the
 phonon-mediated interaction dominates and $g_{\perp}=g_{\parallel}$.}
\label{gamma:fig}
\end{figure}

To solve the gap equation, we approximate the Fermi surface of the
$\gamma$ band as a cylinder with $k_{\rm F}=2\pi/3$ and use the
mean-field approximation.
We have solved the gap equation analytically when the phonon-mediated
interaction
dominates.
The gap in this case is given by
\begin{eqnarray}
&&\psi(k)
\propto
\left\{
\begin{array}{ll}
{\rm const}.&\mbox{for $s$ gap}\\
k_x^2-k_y^2& \mbox{for $d$ gap} \label{sd}
\end{array}
\right.
,
\\
&&\vec{d}(k)\propto(k_x\pm i k_y) \quad\mbox{for $p$ gap}. \label{pgap}
\end{eqnarray}
In Fig.\ref{gamma:fig}, we show the phase diagram in the case where
the phonon-mediated interaction dominates and  $g_{\perp}=g_{\parallel}$.

\noindent
{\it -- Discussions}\\
To conclude, we have studied the superconductive properties of \Ru based on AF
fluctuations and phonon-mediated interactions.
Due to the quasi-one dimensionality,
the odd-parity superconductivity is realized in the $\alpha$ and $\beta$
bands.
The existence of nodes in the gaps of these bands depends
on the symmetry of the system (see Eq.(\ref{nodesym:eqn})) or the
strength of the hybridization of $d_{xz}$ and $d_{yz}$.
The direction of $\vec{d}$ vector is determined by the anisotropy of
the AF fluctuations.

The $\gamma$ band has many possibilities of superconductivity.
If the AF fluctuations direct to the $\hat{c}$ axis and are strong enough,
the odd-parity superconductivity with $\vec{d}\parallel \hat{z}$
is realized, however, if not,  $s$- or $d$-wave superconductivity is
also possible.

There are various possibilities to explain the experimental results of \Ru
~from our theory.
Here we give one of them which is not in the previous theories.
The NMR data \cite{nmr} and the experiment of the Josephson coupling between
In and  \Ru ~\cite{jin} support the odd-parity superconductivity with
$\vec{d}\parallel \hat{z}$.
This is easily realized in our theory with large
$J_{\perp}^{\alpha\beta}(Q)$ and/or $J_{\perp}^{\gamma}(Q,Q)$.
The nodes suggested by NQR\cite{nqr} and specific heat\cite{nishi2} are
likely to be those in the $\alpha$ and $\beta$ bands.
The square vortex lattice observed by the neutron scattering
\cite{riseman} is consistent with the odd-parity superconductivity in
the $\alpha$ and  $\beta$ bands since the superconductivity occurs in
the orthogonal quasi-one dimensional systems.
The $\mu$SR data \cite{muon} can be explained by the
superconductivity in the $\alpha$ and $\beta$ bands with the large
hybridization of $d_{xz}$ and $d_{yz}$ or/and the superconductivity in
$\gamma$ bands with the gap (\ref{pgap}).
Since larger $J_{\perp}^{\alpha\beta}(Q)$ induces larger
$J_{\perp}^{\gamma}(Q,Q)$,  it is likely that the odd-parity
superconductivity in the $\alpha$ and $\beta$ band is followed by that
in $\gamma$ band, so the absence of residual DOS also can be explained.
The transition temperatures could be different between
$\gamma$ and $\alpha$, $\beta$ Fermi surfaces, but there will be a
single $T_{\rm c}$ if the hybridization between
$d_{xy}$ and $d_{yz}$,$d_{xz}$ on the Ru atoms is large enough.

\begin {center}
{\large{\bf Acknowledgments}}
\end {center}
It is a pleasure to thank P. Bourges, Y. Hasegawa, K. Ishida, H.-Y. Kee, Y.
Maeno, K. Maki, Y. Matsuda, Y. Sidis, and M. Sigrist, for  useful discussions.

\end{document}